\documentclass[%
 aip,rsi,amsmath,amssymb,
 reprint,]{revtex4-1}

\usepackage{graphicx}
\usepackage{dcolumn}
\usepackage{bm}
\usepackage{color}

\usepackage{titlesec} 
\titlespacing\section{0pt}{10pt}{4pt}
\titlespacing\subsection{0pt}{10pt}{2pt}

\begin{document}
\title{Variable single-axis magnetic-field generator using permanent magnets}
\author{Oleg Tretiak}
\author{Peter Bl\"umler}
\author{Lykourgos Bougas}
\email{lybougas@uni-mainz.de}
\affiliation{Institut f\"ur Physik, Johannes Gutenberg Universit\"at-Mainz, 55128 Mainz, Germany}

\date{\today}
\begin{abstract}
We present a design for producing precisely adjustable and alternating single-axis magnetic fields based on nested Halbach dipole pairs consisting of permanent magnets only. Our design allows for three dimensional optical and mechanical access to a region with strong adjustable dipolar fields, is compatible with systems operating under vacuum, and does not effectively dissipate heat under normal operational conditions. We present a theoretical analysis of the properties and capabilities of our design, and construct a proof-of-concept prototype. Using our prototype, we demonstrate fields of up to several kilogauss with field homogeneities of better than 5\%, which are harmonically modulated at frequencies of $\sim$1\,Hz with a power consumption of approximately 1\,W. Moreover, we discuss how our design can be modified to generate adjustable quadrupolar magnetic fields with gradients as large as 9.5\,T/m in a region of optical and mechanical access. Our design is scalable and can be constructed to be suitable for table-top experiments, as in the case of polarimetric and magnetometric setups that require strong alternating magnetic fields, but also for large scale applications such as generators.
\end{abstract}

\maketitle

Special arrangements of magnetic structures that increase the magnetic flux on one side of the arrangement while reducing, or even cancelling, it on the opposite side, were first theoretically proposed by Mallinson\,\cite{Mallinson1973}, but later realized by K.\,Halbach\,\cite{halbach1979,halbach1980design,halbach1985application}, who constructed yokeless arrangements of permanent multipole magnets. Nowadays, these arrangements are known as Halbach arrays. \\
\indent Halbach arrays allow for the generation of strong and homogeneous magnetic fields in dipole, quadrupole, or even higher-order multipole configurations. Furthermore, these magnetic fields are the strongest possible per mass of permanent magnet material. Halbach arrays are used in a wide range of applications in science research\,\cite{coey2002}, e.g. particle accelerators\,\cite{halbach1980design}, nuclear magnetic resonance\,\cite{Appelt2006,blumlerNMR,TAYLER2017}, drug targeting\,\cite{nacev2012towards, BlumlerSPIO}, but most importantly in the industrial sector, such as in the cases of brushless motors\,\cite{krishnan2009permanent}, magnetic gears\,\cite{jian2010coaxial}, magnetic refirgeration\,\cite{Trevizoli2015}, and magnetically levitating trains\,\cite{maglev}, and most notably in ``kitchen-refrigerator'' and pin-board magnets. \\
\indent Recent developments in polarimetry\,\cite{Sofikitis2014,Bougas2015} and magnetometry \,\cite{Wickenbrock2016,Zheng2017,Chatzidrosos2017} require the use of strong and adjustable, or alternating, magnetic fields, but the ability to maintain three-dimensional (3D) optical and mechanical access in these applications is, in most cases, critical. A typical solution is to use electromagnets, but their usage can be undesirable due to power-supply-related noise\,\cite{Sofikitis2014,Bougas2015,Wickenbrock2016,Zheng2017,Chatzidrosos2017}, or untenable due to power-availability constraints. Moreover, in the case of high-field generation using electromagnets, active cooling typically becomes necessary, introducing additional mechanical noise into the system\,\cite{Zheng2017}. Leupold et al.\,\cite{Leupold1988} presented the idea of nested, concentric, and rotatable Halbach cylinders for the generation of adjustable, homogeneous magnetic fields using permanent magnets, circumventing issues associated with the use of electromagnets, but these, and similar presented systems, lack the 3D optical and mechanical accessibility. \\
\begin{figure}[h]
		\includegraphics[width=0.8\linewidth]{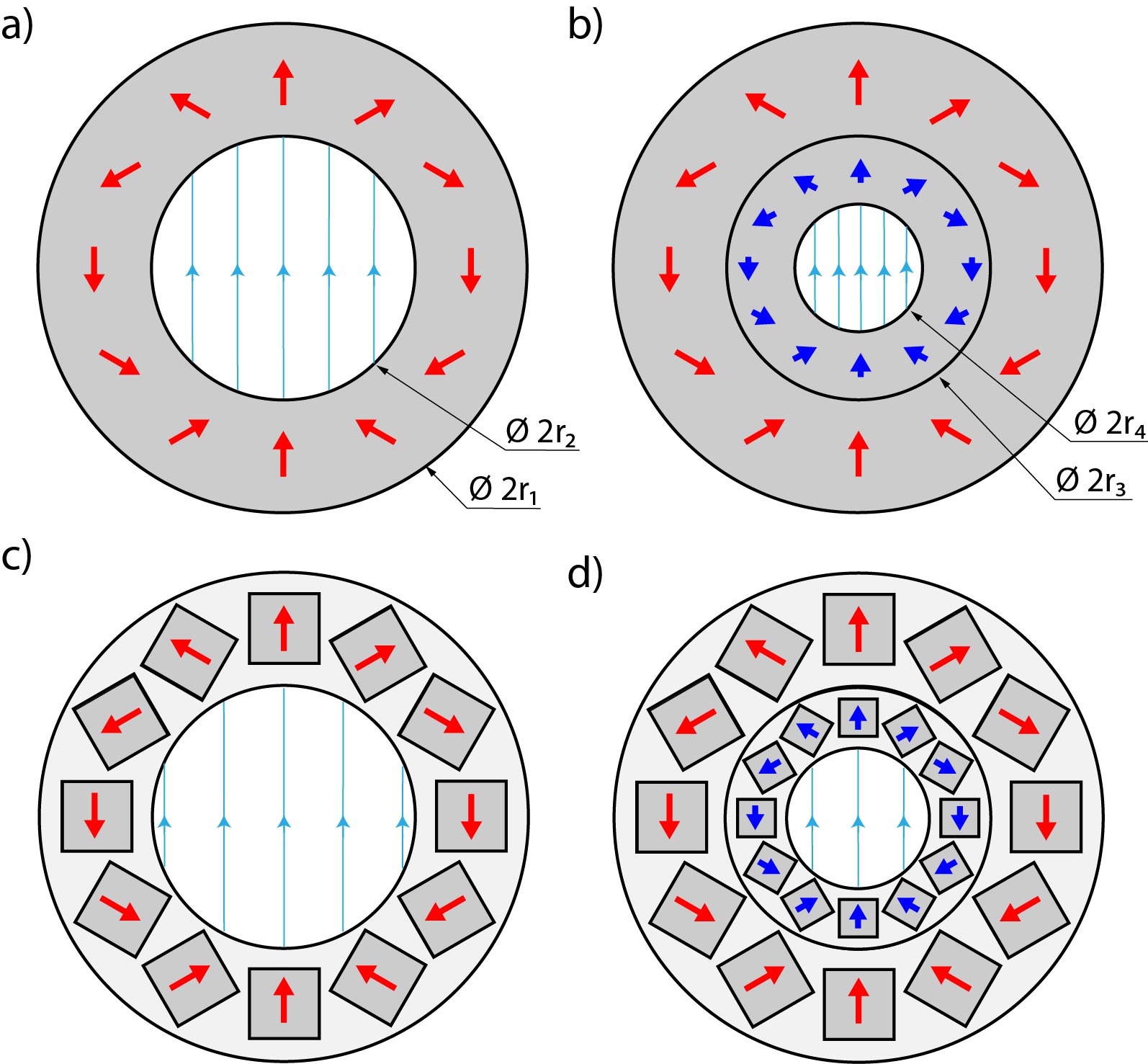}
	\caption{Schematic illustration of ideal cylindrical Halbach dipoles: (a) single (b) nested, and schematic illustration of discretized cylindrical Halbach dipoles: (c) single (d) nested. The magnetization of each structure is represented by the red and blue arrows - in the discretized case, the arrows denote the magnetization orientation of cuboid magnets. The solid cyan lines show the homogeneous dipole magnetic fields obtained within the plane of the Halbach dipoles (the line density schematically illustrates the strength of the field obtained for each configuration).}	
    	\label{fig:fig1}
\end{figure}
\indent In this article, we present a design based on concentrically nested Halbach dipole pairs, axially separated by an appropriate distance. This particular arrangement allows for the production of single-axis adjustable and alternating magnetic fields at the central plane of the structure while enabling 3D optical and mechanical access to a region of a homogeneous magnetic field. \\
\indent We first describe the principles of operation of our design using a theoretical analysis. Then, we present a proof-of-concept prototype along with magnetic field measurements obtained with it, demonstrating the capabilities of our system. Finally, we discuss the ability to create quadrupolar magnetic fields using our design, and present simulations for the attainable magnetic field gradients. Our primary motivation is the employment of the proposed system in cavity-enhanced polarimetry\,\cite{Sofikitis2014,Bougas2015}, where the generation of adjustable and reversible circular birefringent effects (e.g. Faraday rotation) is typically required. We also discuss the flexibility of our device and several possible applications that potentially benefit from it.
 \begin{figure}[h]
		\includegraphics[width=\linewidth]{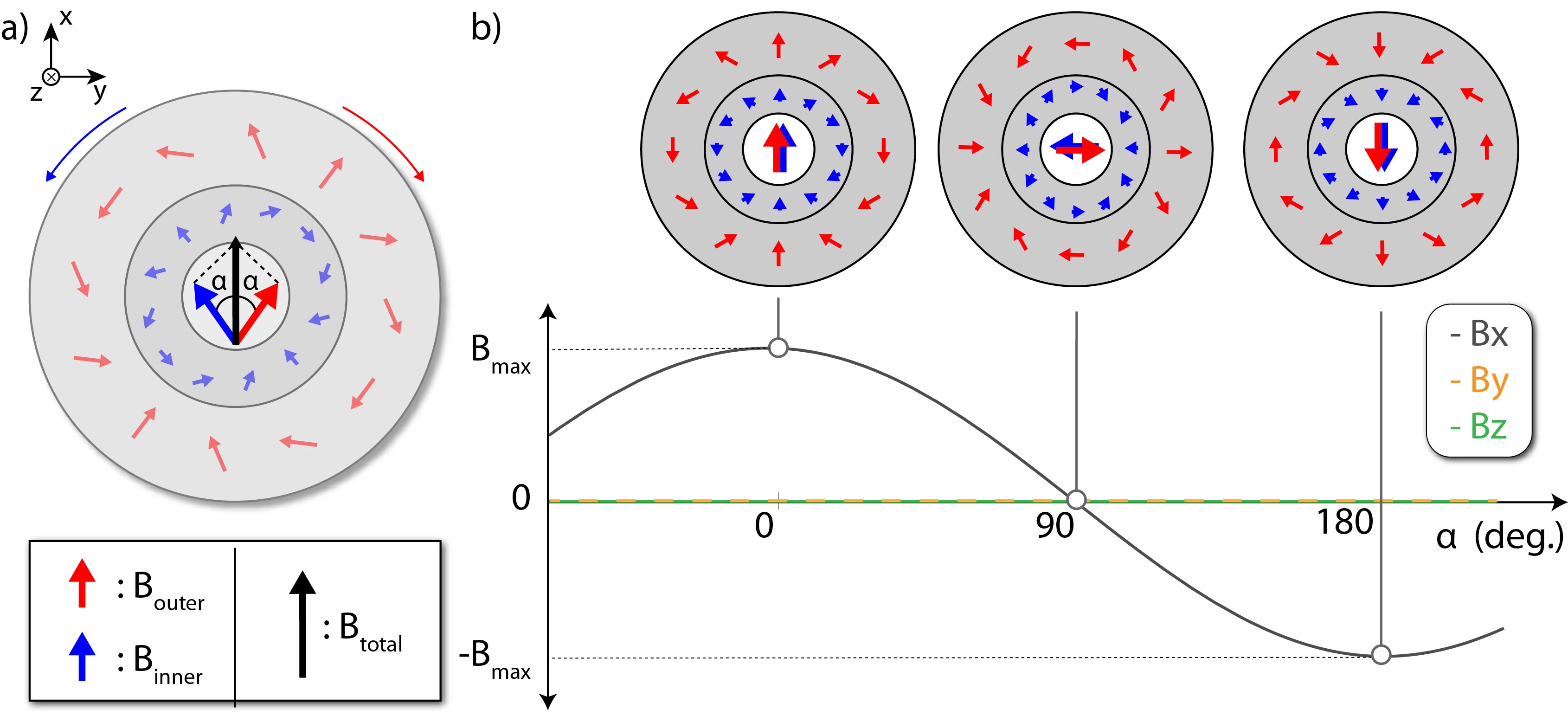}
	\caption{a) Schematic illustration of the magnetic field generated within the nested cylindrical ideal Halbach dipoles as a function of their mutual angle. The magnetization for the outer and inner cylindrical dipole is represented by the red and blue arrows, respectively. b) By continuously adjusting the mutual angle of the nested dipoles (in a counter-rotating fashion), one can generate an oscillating homogeneous single-axis magnetic field.}
    	\label{fig:fig2}
\end{figure}
\section{\label{sec:Theory}Theory of operation}
\subsection{Ideal Halbach dipoles: single and nested}
Here we focus on cylindrical Halbach arrays that generate homogeneous dipolar fields, i.e. Halbach dipoles, as shown in Fig.\,\ref{fig:fig1}. An ideal Halbach dipole, represented as an infinitely long tube with continuously rotating magnetization inside the material of the wall, produces a homogeneous magnetic field within a cross-sectional plane of strength ${\mathbf{B}_{\rm{outer}}=\mathbf{B}_{\rm{rem}} \cdot \ln \left ( r_{1}/r_{2} \right )}$, where $\mathbf{B}_{\rm{rem}}$ is the remanence of the permanent magnetic material, and $r_{1}$ and $r_{2}$ are the outer and inner radii of the tube, respectively [see Fig.\,\ref{fig:fig1}\,(a)]\,\cite{halbach1980design}.\\
\indent Let us now assume a configuration of two concentric, \textit{nested}, Halbach dipoles as shown in Fig.\,1\,(b). In this case, the inner dipole now produces a homogeneous magnetic field of strength ${\mathbf{B}_{\rm{inner}}=\mathbf{B}_{\rm{rem}} \cdot \ln \left ( r_{3}/r_{4} \right )}$. If this inner dipole has an inner and outer radius of:
\begin{equation}
r_3=r_2 \quad\rm{and}\quad r_4 = r_2^2/r_1,
\label{eq:r4}
\end{equation}
respectively, then $|\mathbf{B}_{\rm{outer}}| = |\mathbf{B}_{\rm{inner}}|$. Since the resulting field in the center of the dipole pair is given by the vector addition of $\mathbf{B}_{outer}$ and $\mathbf{B}_{inner}$, i.e. $\mathbf{B_{\rm{tot}}}=\mathbf{B}_{\rm{outer}}+\mathbf{B}_{\rm{inner}}$, it becomes apparent that for this specific geometry, appropriately equal and opposite rotations of the two nested dipoles yields any desired field in the range of $\pm |\mathbf{B}|^{\rm{max.}}_{\rm{tot}}$ along one axis of the system, as schematically depicted in Fig.\,\ref{fig:fig2}. Such arrangements of nested Halbach dipoles allow for great versatility in choice of operating alternating frequency as well as in field tuning\,\cite{Leupold1988,blumlerNMR}. While this design allows for single-axis variable magnetic fields, it does not allow for flexible 3D optical and mechanical access.\\
\subsection{Discretized Halbach dipole}
To translate the idea of nested Halbach dipoles in a realistic design we need to modify the ideal model presented above, since an infinitely long tube of continuously changing magnetized material is impossible to realize. Various discrete versions of the ideal case have been suggested to approximate it\,\cite{blumlerNMR}, with the most straightforward discretization being the division of the cylindrical structure into segments of identical shape and magnetization, but with different magnetization directions\,\cite{Raich2004,Hills2005}. \\
\indent In this work we choose to focus on cylindrical arrays of square-shaped permanent magnets as the discrete equivalent of ideal Halbach dipoles, similar to the ones depicted in Fig.\,1\,(c) (representing a single Halbach dipole), and (d) (representing nested Halbach dipoles). For the single discretized Halbach dipole [Fig.\,1\,(c)], we can estimate the resulting magnetic field within and outside its magnetization plane (namely the x-y plane; see Fig.\,\ref{fig:fig2}). The simplest approach is to model the cylindrical array of the $N$ square-shaped permanent magnets as an array of $N$ point-dipoles. In this case, the resultant magnetic field ${\mathbf {B_{array}}}$, at a particular point in space with radius vector $\mathbf{R}$, is the vector sum of the magnetic field $\mathbf {B_i}$ of each of the $N$ dipoles at positions $r_i$\,\cite{soltner2010dipolar}:
\begin{equation}
	{\mathbf {B_{array}}} ({\mathbf R})= \sum_{i=1}^{N} {\mathbf {B_i}}({\mathbf R- \mathbf r_i})\!.
\label{eq:B}
\end{equation}
By using this approximation we can estimate the strength of the magnetic field of the discretized Halbach dipole along its x-axis as a function of the distance from its central plane (i.e. along the z-axis, Fig.\,\ref{fig:fig2}) as follows\,\cite{soltner2010dipolar}:
\begin{equation}
	B_x(x=0, y=0, z) = \frac {3}{8} N \mu_0\frac{m\,r^2}{ \pi  \left(r^2+z^2\right)^{5/2}},
    \label{formulaBy}
\end{equation}
where $\mu_0 = 4\pi \cdot 10^{-7}\,\rm{T}\cdot \rm{m/A}$ is the permeability of vaccum, $m$ is the magnetic dipole moment of a single dipole (we assume here that each dipole, representing each magnet, has the same magnetic dipole moment), $r$ the effective radius of the array of the point dipoles, and $z$ the distance from the array's x-y plane.\\
\indent By appropriately choosing the characteristics, dimensions, and orientation of the two arrays comprising a nested Halbach dipole, we can carefully tune the strength and homogeneity of the dipolar field inside and outside the plane of the nested dipoles. It is this property that we exploit for the creation of a device that allows for adjustable, homogeneous, magnetic fields within a region of 3D optical and mechanical access, by using concentrically nested Halbach dipole pairs, axially separated by an appropriate distance.
 \section{Proof-of-concept design}
\subsection{Nested Halbach dipole pairs}
\begin{figure}[!h]
	\centering
		\includegraphics[width=\linewidth]{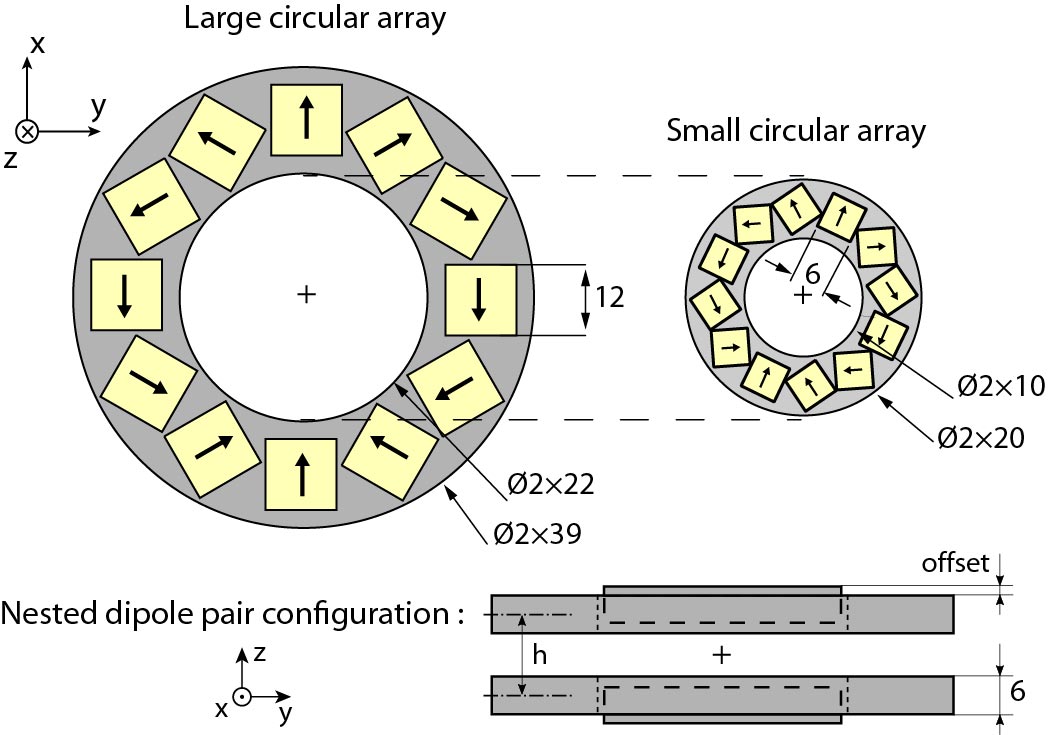}
			\caption{Schematic of the discretized Halbach arrays used in our proof-of-concept prototype device, which consists of concentrically nested Halbach dipole pairs. We use a total of four cylindrical arrays for our design: two identical large (outer) and two identical small (inner) ones; using one of each we create a nested Halbach dipole pair. Each array consists of 12 cubic neodymium magnets (yellow) which are encased in a plastic housing (gray). In the nested Halbach dipole pair configuration, we define as $h$ the distance between the central planes of the nested dipoles, and as offset the distance between the central planes of the small (inner) and large (outer) arrays in each nested configuration. By tuning $h$ and the offset, one can tune the magnetic field strength at the geometrical center of the structure (marked by a cross symbol), which is the area of 3D optical and mechanical access. All dimensions are in mm.}
	\label{fig:sizes}
\end{figure}

For a prototype device that integrates the above discussed concepts and is appropriate for our applications, we use two identical pairs of nested Halbach cylindrical arrays, shown in Fig.\,\ref{fig:sizes}. Note that we choose the dimensions of the cylindrical structures to be in accordance with the design criteria imposed by Eq.\,\ref{eq:r4}.\\
\indent Each cylindrical array consists of 12 magnets, and we use polyoxymethylene (POM) as a material for the cylindrical support structures to encase the magnets. We use cuboid-shaped neodymium (Nd$_2$Fe$_{14}$B) permanent magnets (N50 alloy; B$_{\rm{rem}}\approx1.4$\,T) of two different sizes: $6\times6\times6$\,mm$^3$ for the inner arrays, and $12\times12\times6$\,mm$^3$ for the outer arrays. The thickness of the plastic housings is chosen to be equal to the height of the magnets (i.e. 6\,mm). We characterize the magnets separately, and find their magnetization to vary no more than $\sim$5\%. We note here that great care is required when assembling systems consisting of strong permanent magnets as contact can strongly influence their magnetic properties.\\
\begin{figure}[!h]
	\centering
		\includegraphics[width=\linewidth]{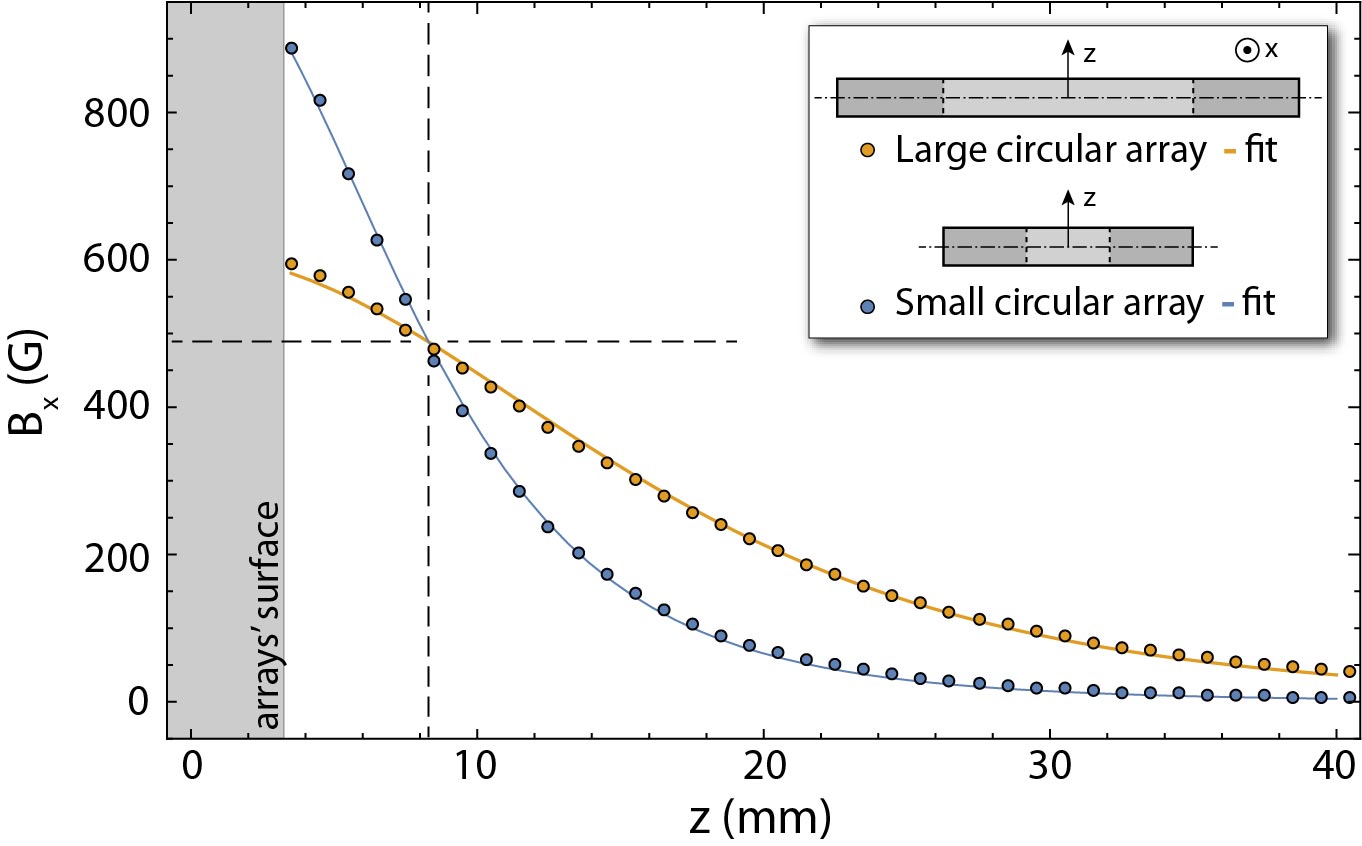}
			\caption{Measurements of the magnetic field strength along x-axis for the large (orange) and small (blue) constructed Halbach arrays as a function of distance from their central planes. The arrays' dimensions are provided in Fig.\,\ref{fig:sizes}. Solid lines are fits to the data using Eq.\,\ref{formulaBy}. Dashed lines indicate the crossing point (in the z-axis) where the fields from each array become equal. }
	\label{fig:ydecays}
\end{figure}
\begin{figure*}[ht]
	\centering
		\includegraphics[width=\linewidth]{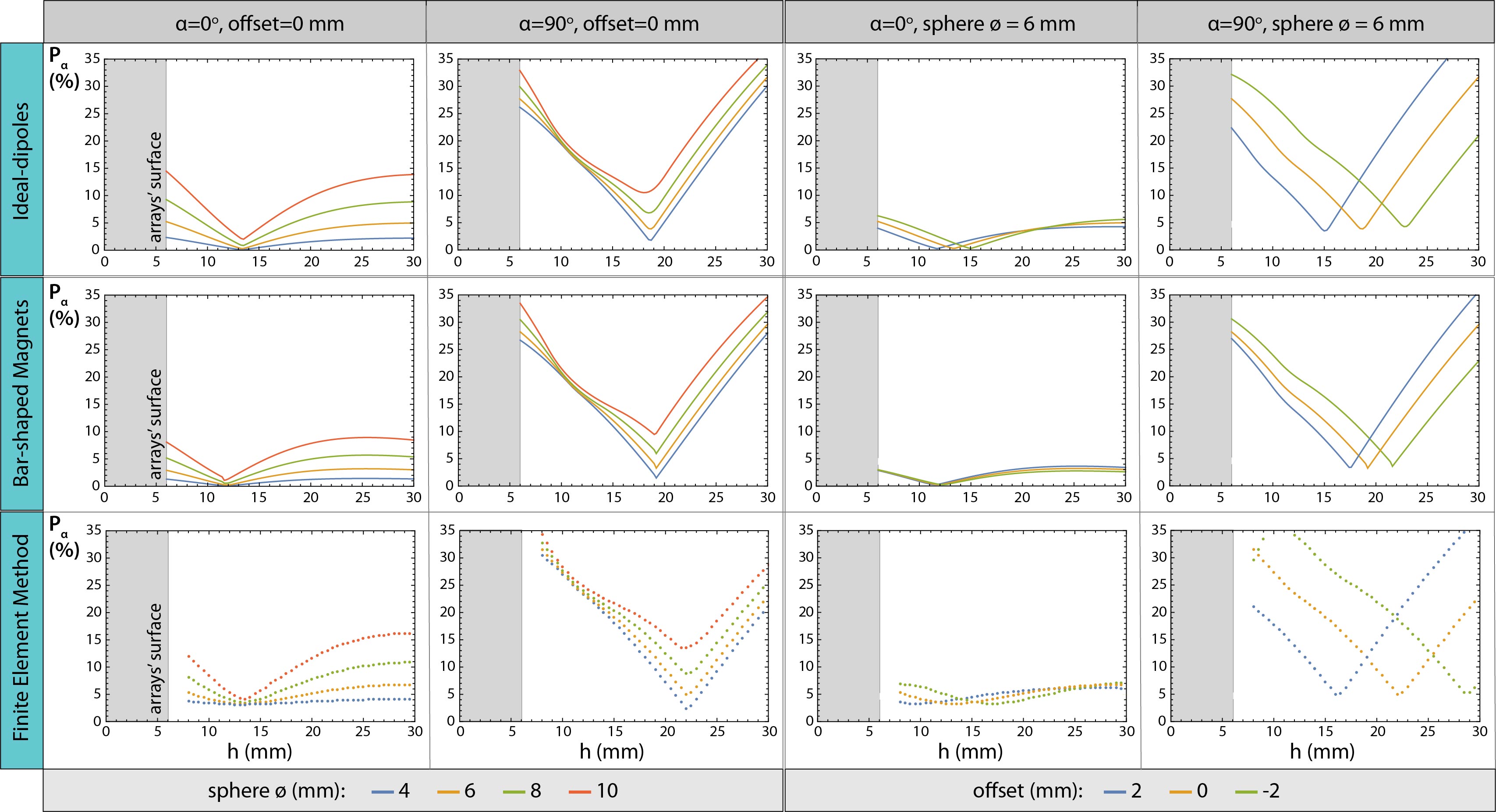}
			\caption{Theoretical estimation of the $P_{\alpha}(x,y,z)$ parameter for different design parameters. See text for more details.}	
	\label{fig:homo}
\end{figure*} 
\begin{figure}[ht]
	\centering
		\includegraphics[width=0.4\textwidth]{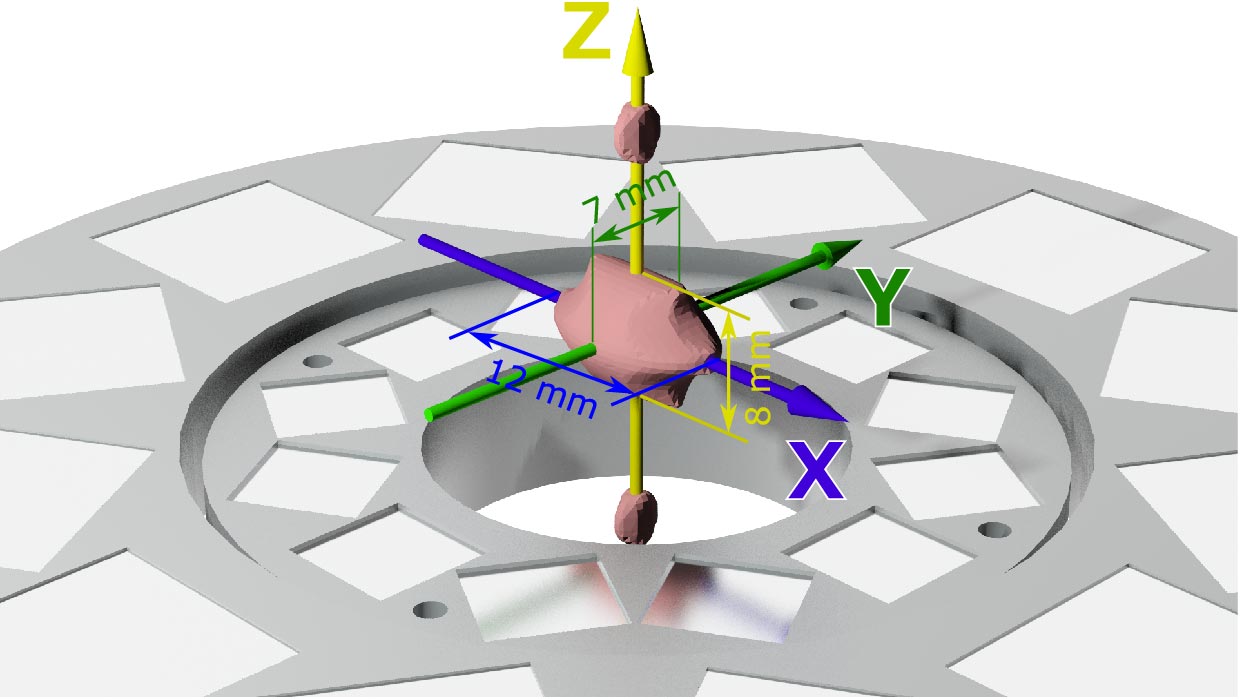}
			\caption{Theoretical simulation of the exact volume (light red) within which the $P_{\alpha}(x,y,z)$--parameter and the B$_x$ magnetic field strength vary less than 5\%, for a gap of $h\!=\!17$\,mm and an offset of $+2$\,mm (Fig.\,\ref{fig:sizes}). This volume is 3D accessible optically and mechanically. The lower nested Halbach arrays are shown as a size reference, while the upper arrays are not shown.}
	\label{fig:5per-zone}
\end{figure}
Our goal is to obtain strong, homogeneous, magnetic fields along one axis of the system (x-axis; Fig.\,\ref{fig:sizes}) in an area with optical and mechanical axis (with its geometrical center marked by a cross in Fig.\,\ref{fig:sizes}), that can be adjustable and reversible, while minimising the strength of the magnetic field along the other axes. For this reason, we start by measuring the magnetic field strengths of the built arrays along the x-axis as a function of displacement from their central planes (i.e. along the z-axis; Fig.\,\ref{fig:sizes}). In Fig.\,\ref{fig:ydecays} we present our measurements which we analyze using Eq.\,\ref{formulaBy}, yielding the effective magnetic moments for the large and small magnets to be $0.161(1)$\,A$\cdot$m$^2$ and $0.721(7)$\,A$\cdot$m$^2$, respectively, and their radial position around the arrays to be located at $r=14.1(1)$\,mm and $r=27.8(1)$\,mm, respectively. These results are consistent with the parameters and geometry of our design.\\
\indent Most importantly, we see that there exists a position outside the arrays (along the z-axis), where the strengths of the B$_x$ magnetic field component produced by the large and small cylindrical arrays are equal. Therefore, as the nested dipoles are rotated in opposite directions, we expect that at this intersection point the B$_x$ magnetic field component will cancel when the angle between the nested arrays of $90^\circ$, and completely reverse for $180^\circ$ (Fig.\,\ref{fig:fig2}). It is at this position that the nested dipole Halbach pair configuration allows for single-axis variable magnetic fields with 3D optical and mechanical access. Moreover, this intersection point indicates that the gap $h$ (\,Fig.\,\ref{fig:sizes}), between the axially separated nested arrays, must be approximately equal to twice this intersection distance. For the built arrays we use in our prototype device, the separation between them should be, therefore, set to $h\simeq\!16\!-\!18$\,mm. We note here that the introduction of an additional offset between the central planes of the small (inner) and large (outer) nested arrays (see Fig.~\ref{fig:sizes}), allows us to further tune the intersection point and, as a consequence, the strength of the total magnetic field strength at that point. This offset adds an extra degree of freedom in the design parameters of our proposed system.\\
\indent To decide on the final design parameters for our prototype device, we use the geometric characteristics of our built arrays (Fig.\,\ref{fig:sizes}) and the results shown in Fig.\,\ref{fig:ydecays} to theoretically estimate the attainable magnetic field strength and homogeneity for possible configurations of our device in the area between the axially separated nested dipole pairs. In particular, as a measure of the maximally attainable magnetic field homogeneity in a given geometry we use the following parameter:
\begin{equation}
P_{\alpha}(x,y,z) = \max\limits_{\{x,y,z\}\in S}\frac{|\rm{B}_{yz}(\alpha,x,y,z)|}{|\rm{B}_x(0,0,0,0)|}  \cdot 100\%,
\end{equation}
where $\alpha$ is the angle between the nested arrays (Fig.\,\ref{fig:fig2}), $|\rm{B}_{x}(0,0,0,0)|$ is the absolute value of B$_{x}$ on the x-axis at the center of the Halbach array system for $\alpha =0^\circ$ (equal to the case of $\alpha=180^\circ$), B$_{yz}(\alpha,x,y,z)$ is the projection of the magnetic-density flux vector on the yz-plane at a point with coordinates $\{x,y,z\}$, and $S$ is the volume around the $\{x,y,z\}$ point within which we estimate the $P_{\alpha}$ parameter. We use three different approaches to calculate the magnetic field strengths and their homogeneity - as characterized by the $P_{\alpha}(x,y,z)$ parameter - for a given design geometry of the proposed nested Halbach-pair device: we use (a) a dipole model, where each permanent magnet in the arrays is modelled as an ideal magnetic dipole; (b) an analytical model that uses analytical expressions for the magnetic field of a cuboid-shaped permanent magnet following the work of Ref.\,\cite{Herbert2005}; and (c) a COMSOL-based finite element method that uses as inputs the geometry of our design and the measured magnetic moments of our magnets. \\
\indent In Fig.\,\ref{fig:homo} we show the dependence of the $P_{\alpha}(x,y,z)$-parameter on the distance between the central planes of the Halbach arrays for two extreme orientations, parallel ($\alpha=0^{\circ}$) and perpendicular ($\alpha=90^{\circ}$), and for different offset parameters. In particular, we estimate the value of the $P_{\alpha}(x,y,z)$--parameter within a sphere ($S$) of a given diameter centered around the geometrical center of the structure, estimating thus the homogeneity within a given volume for the structures shown in Fig.\,\ref{fig:sizes}. We see, that, for $\alpha=0^{\circ}$ the magnetic field homogeneity is better than 5\% within a sphere of $\approx8$\,mm in diameter and for dipole separation gaps in the range of $h\approx10-20$\,mm. Moreover, for $\alpha=90^{\circ}$ we see that similar homogeneities are attainable, suggesting that for carefully selected gaps we can achieve overall field homogeneities of better than 5\% while the system is rotating. We perform the same calculations assuming an additional offset (Fig.\,\ref{fig:sizes}). We see that by adjusting this offset we can tune the optimum gap distance for the system without affecting its field homogeneity, ensuring thus that one can add a small offset in the final design to allow for an increase in the maximum attainable magnetic field without significantly affecting its overall performance. Additionally, in Fig.\,\ref{fig:5per-zone} we show the exact shape of the volume within which two conditions are satisfied simultaneously: a) $P_{\alpha}(x,y,z)$ is less than 5\% for any angle $\alpha$, and b) the strength of B$_x$ does not vary more than 5\% for $\alpha\!=\!0^{\circ}$. For these calculations we use a gap of $h\!=\!17$\,mm. This volume and its predicted homogeneity are sufficient for our polarimetric and magnetometric applications, but further adjustment is possible for suitability in other applications. In the following section we present a proof-of-concept prototype device, and, for its construction, we use the presented numerical simulations as our guide.
\subsection{Instrument Design}
\begin{figure}[!t]
	\centering
		\includegraphics[width=0.4\textwidth]{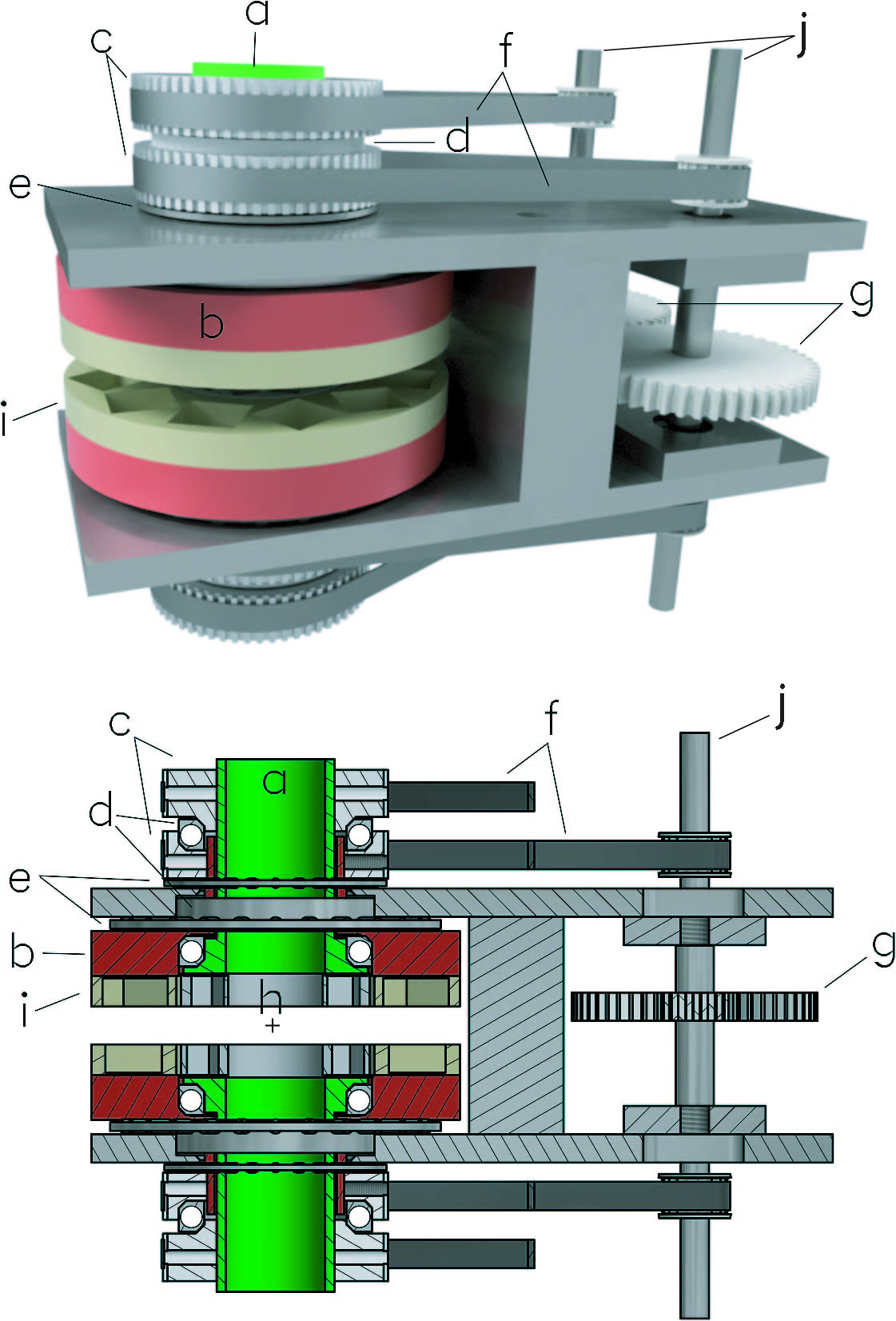}
			\caption{a) Rendering and b) schematic full sectional side view of the complete assembled system that consists of: the Halbach arrays [inner (h) and outer (i) arrays], and the rotation mechanism and support structure [(a)-(f)]. See the text for additional details. The geometrical center of the structure is marked by a cross.}	
	\label{fig:setup-section-view}
\end{figure}

\indent In Fig.\,\ref{fig:setup-section-view} we present a rendering and a schematic cross-sectional side view of our assembled prototype device. The inner and outer Halbach cylindrical arrays [with dimensions and characteristics as presented in Fig.\,\ref{fig:sizes}, and annotated as (h) and \,(i) in Fig.\,\ref{fig:setup-section-view}], are screwed on a rotating system which consists of a hollow stainless-steel-based hub assembly [(a) and (b)], constrained axially and radially by angular ball and thrust roller bearings [(d)-(e)]. This hollow hub assembly allows for the insertion of, for example, optomechanical mounts enabling the positioning of optics within the region of strong, homogeneous magnetic fields. The hub assembly together with the bearings are pressed and held in position by pulleys (c) on a stainless-steel plate. An external thread on the hub assembly (a) also enables the use of a locking mechanism, e.g. a locking nut (external thread not shown in Fig.\,\ref{fig:setup-section-view}). The same mechanism is mirrored to host and control the rotation of the opposite nested Halbach dipole pair, and is held in position by the overall stainless steel body of the device. Finally, a system of timing belts (f) and gears (g) allows for the synchronized counter-rotation of the inner (h) and outer (i) Halbach arrays. The whole system can be driven using the gear shafts (j), manually and/or electrically by an external motor. This prototype design is adaptable to produce strong magnetic fields in any direction in a horizontal plane, and is suitable for operation under vacuum. We wish to emphasize that the use of stainless steel in our prototype is chosen for reasons of robustness and it can affect the magnetic properties of the design, but other materials can be used without obvious shortcomings. \\
\indent The presented design is scalable, but for suitability to our experimental plans, and in accordance with our simulations, we have chosen a gap of $h\simeq17$\,mm and an offset of $+2$\,mm for our prototype device.
%
%
%
%
 \section{Experimental measurements}
 \begin{figure}[!h]
		\includegraphics[width=0.9\linewidth]{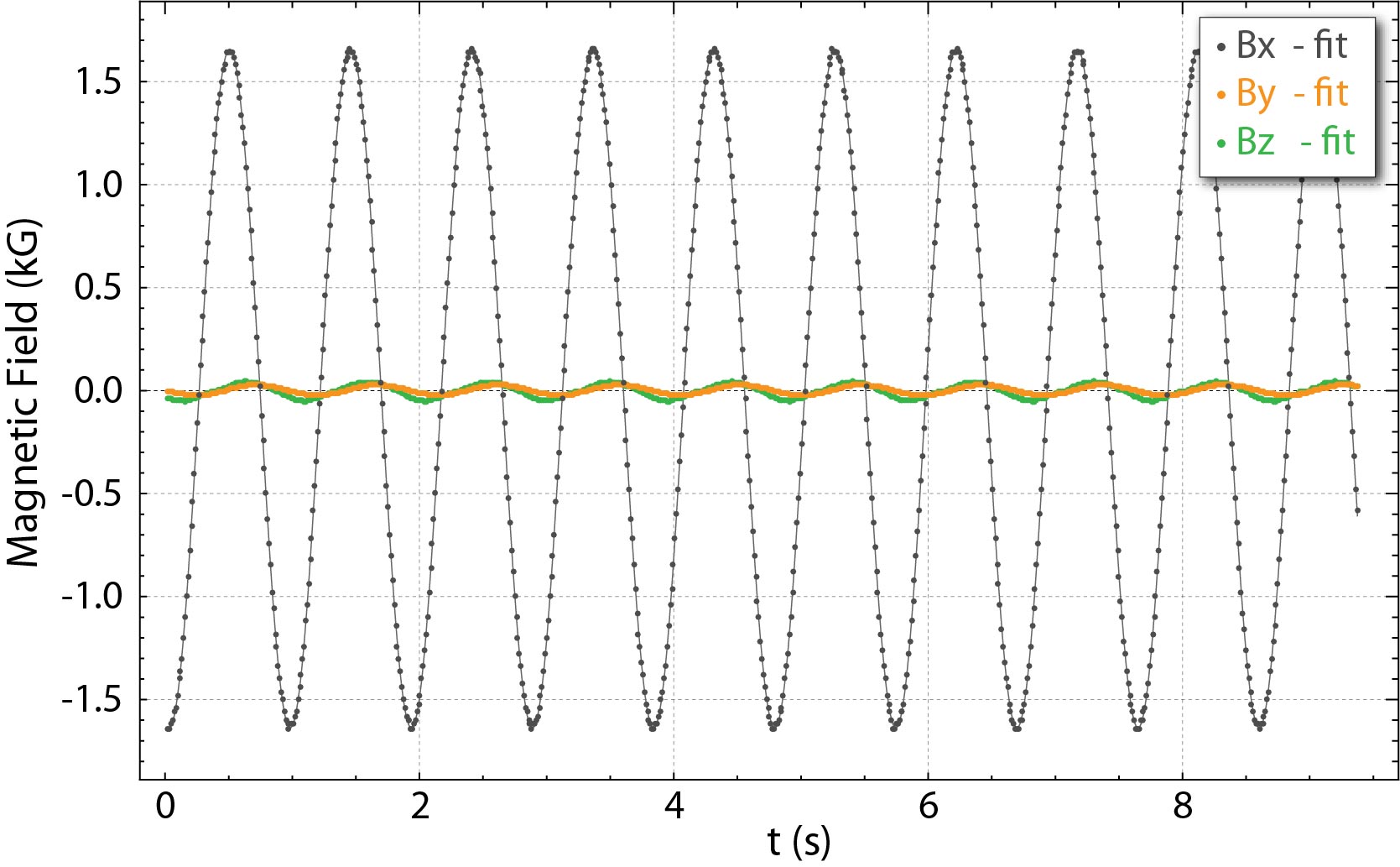}
			\caption{Magnetic field measurements using our prototype device at its geometrical center (marked by a cross; Fig.\,\ref{fig:setup-section-view}) while rotating the system using a stepping motor. Black points-- B$_x$; yellow points -- B$_y$; green points -- B$_z$; solid lines -- harmonic fits of B$_{x,y,z}$.}
	\label{fig:oscillations}
\end{figure}
 \begin{figure*}[!t]
	\centering
		\includegraphics[width=0.9\textwidth]{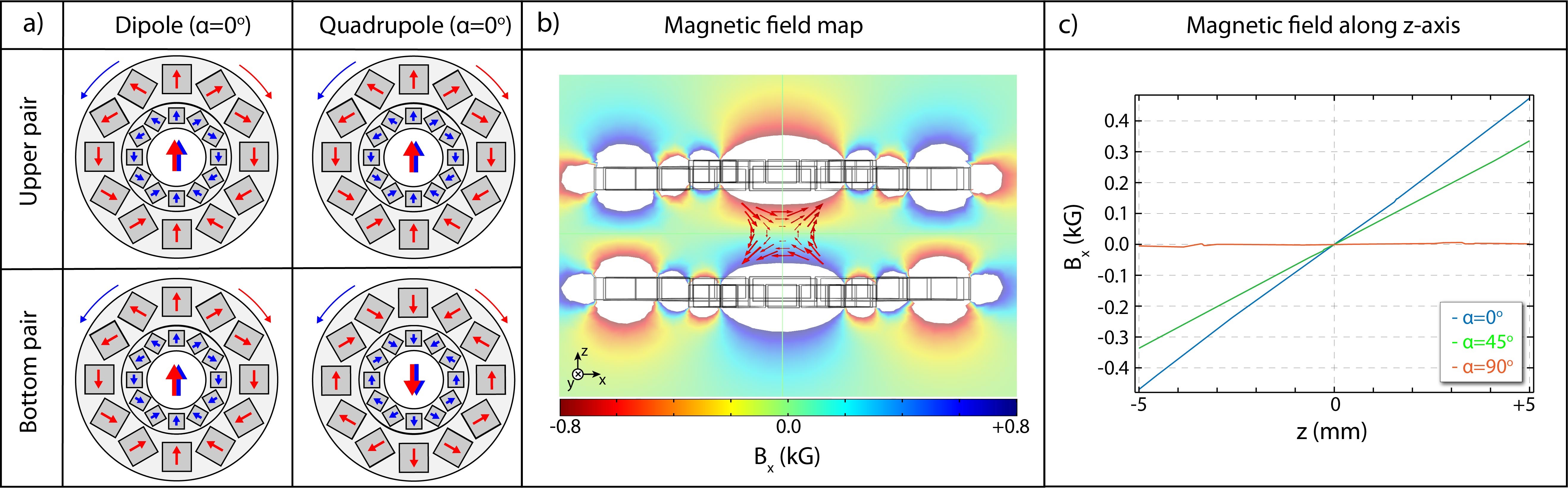}
			\caption{Generation of quadrupolar magnetic fields using axially separated concentrically nested Halbach dipole pairs. a) Schematic representation of the initial orientation parameters for the axially separated arrays for the generation of dipole vs. quadrupolar fields. b) Using a finite element method (COMSOL) we simulate the magnetic field B$_x$ component in xz-plane generated by our design. Red arrows show the directions and relative magnitude of the magnetic flux density vector $\mathbf{B}$. c) Magnetic field gradient ($^{\rm{dB}_x}/_{\rm{dz}}$) obtained using our design. For these simulations we assume a gap of 30\,mm and an offset of +2\,mm (Fig.\,\ref{fig:sizes}).}
	\label{fig:quadrupole}
\end{figure*}
  \subsection{Adjustable dipole magnetic fields}
In Fig.\,\ref{fig:oscillations} we present measurements of the magnetic field generated at the geometrical center of our prototype design (marked by cross; Fig.\,\ref{fig:setup-section-view}). For our measurements we use a transverse and an axial Hall probe magnetometer (HIRST Magnetic Instruments GM08 magnetometer) with a $<1$\,mm sensor size. We see that with our prototype system we can obtain fields as strong as $\sim$1.6\,kG at the geometrical center, where we have 3D optical and mechanical access. We observe that any remnant field strengths along the other two axes (y\,\&\,z) to be less than 4$\%$ of the maximum attainable field strength along the x-axis. Moreover, with the use of a stepping motor (Sanyo Denki 103H7123) we are able to continuously rotate our system at a rate of $\sim$1\,Hz (we are limited by our stepping motor's holding torque, which is frequency dependent). For this rotational frequency, and for a torque required to rotate the system of approximately 0.2\,N$\cdot$m, we estimate the power consumption for generating this alternating field to be $\sim$1\,W. We also perform a harmonic fit of our measurements and verify that our system has a stable and harmonic behaviour. Most critically, we observe that the remnant fields are out of phase with respect to the magnetic field along the x-axis, which suggests that by finely adjusting the geometry parameters of our design we can tune the off-axis fields to be completely out of phase with the field along x-axis. Finally, we note here, that, if an anharmonic magnetic-field oscillation is desirable, it can be achieved by simply modulating the speed of the driving motor.
\subsection{Adjustable quadrupole magnetic fields}
An additional feature of our design is the ability to generate homogeneous magnetic fields of multipole configurations. In particular, quadrupole magnetic fields can be generated inside single Halbach cylinders with quadrupolar magnetization distributions\,\cite{blumler2016proposal}, but these cannot be adjusted and single arrays do not allow for 3D optical and mechanical access.\\
\indent Using our design and by oppositely aligning the axially separated nested Halbach pairs, as schematically presented in Fig.\,\ref{fig:quadrupole}\,(a), we can create an adjustable quadrupolar magnetic field along the x-z plane. In particular, in Fig.\,\ref{fig:quadrupole}\,(b) we present a vector field map of the quadrupole magnetic field attainable using our design, and in Fig.\,\ref{fig:quadrupole}\,(c) we present the attainable magnetic field gradient. For these simulations we choose a gap $h$ (Fig.\,\ref{fig:sizes}) equal to 30\,mm for which we obtain a permanent quadrupolar field along the x-z plane for any angle $\alpha$. Moreover, we see that we can obtain magnetic field gradients ($^{\rm{dB}_x}/_{\rm{dz}}$) of up to 9.5\,T/m, and by adjusting the relative orientation of the axially spaced nested Halbach pairs we can adjust this gradient, a unique feature of our design. The presented configuration could be used in fields where linear magnetic gradients are required, e.g.: nuclear magnetic resonance imaging\,\cite{blumler2016proposal}, ion-beam focusing\,\cite{qin1991construction}, spin-polarized atom trapping\,\cite{PhysRevLett.59.672}. Detailed investigations of generation of quadrupolar fields using our design will be part of future work.

 \section{Conclusions}
In conclusion, we present here a design for producing variable single-axis magnetic fields based on nested Halbach dipole pairs consisting of permanent magnets. We provide the design parameters for a prototype device with moderate dimensions that allows for 3D optical and mechanical access to a region with an adjustable field of up to several kilogauss and with field homogeneities of better than 5\%. Our device is compatible with systems operating under vacuum and does not effectively dissipate heat under normal operational conditions. Moreover, using our prototype, we demonstrate harmonic generation of $\sim$1.6\,kG strong magnetic fields at a frequency of $\sim1$\,Hz, with an estimated power consumption of $\sim$1\,W. We finally discuss the ability to create adjustable quadruple magnetic fields, making our design appropriate for other applications such as utility in linear accelerators and in magnetic particle trapping applications. \\
\indent While the design parameters of our prototype device were selected to be suitable for table-top scientific experiments such as highly sensitive polarimeters and magnetometers built in our laboratories, it is important to emphasize that the proposed design can be scaled in all dimensions, with particular interest in industrial applications, such as DC motor generators.

\section*{Acknowledgements}
We would like to thank Prof. D. Budker for his constant support and help during this work. LB thanks G. Iwata and A. Lev for fruitful discussions. This research was supported by European Commission Horizon 2020 (grant no. FETOPEN-737071), ULTRACHIRAL Project. LB acknowledges support by a Marie Curie Individual Fellowship within the second Horizon 2020 Work Programme.

\bibliography{Halabachbib}
\end{document}